# Surface Plasmon-Enhanced X-ray Ultraviolet Nonlinear Interactions

H. Aknin[1], O. Sefi[1], D. Borodin[1], J.-P. Rueff[2,3], J. M. Ablett[2], and S. Shwartz [1*]

[1]Physics Department and Institute of Nanotechnology, Bar-Ilan University, Ramat Gan, 52900 Israel

[2]Synchrotron SOLEIL, L'Orme des Merisiers, Départementale 128, 91190 Saint-Aubin, France

[3]Sorbonne Université, CNRS, Laboratoire de Chimie Physique-Matière et Rayonnement, LCPMR, F-75005 Paris, France

X-ray–matter interactions are intrinsically weak, and the high energy and momentum of X-rays pose significant challenges to applying strong light–matter coupling techniques that are highly effective at longer wavelengths for controlling and manipulating radiation. Techniques such as enhanced coupling between light and electrons at a metal-dielectric interface or within nanostructures, as well as the Purcell effect—where spontaneous emission is amplified near a metallic surface—are not applicable to X-rays due to their fundamentally different energy and momentum scales. Here we present a novel approach for coupling X-rays to surface plasmon polaritons (SPPs) by entangling X-ray photons with SPPs in the ultraviolet (UV) range through X-ray-to-UV spontaneous parametric down-conversion (SPDC) in aluminum. The distinct characteristics of the SPPs are imprinted onto the angular and energy dependence of the detected X-ray photons, as demonstrated in this work. Our results highlight the potential to control X-rays using SPPs, unlocking exciting opportunities to enhance X-ray–matter interactions and explore plasmonic phenomena with atomic-scale resolution—a capability uniquely enabled by X-rays.

## Main

Surface plasmon polaritons (SPPs) are surface waves that propagate along the interface between a metal and a dielectric[1,2]. They have attracted significant attention due to their ability to confine and enhance electromagnetic fields at this boundary[3,4]. Since their discovery, SPPs have been harnessed to control and enhance a wide range of linear and nonlinear optical phenomena[4–6]. Notable examples include the enhancement of spontaneous emission from quantum emitters[7–12] and the amplification of nonlinear optical interactions[12–17]. The spectral and directional properties of SPPs enable strong capability to manipulate light at material interfaces, leading to the development of groundbreaking optical devices[18–25]. Furthermore, the enhanced interactions facilitated by SPPs have enabled the advancement of techniques such as super-resolution imaging[26–28] and highly sensitive spectroscopy[26,29–32].

While the field of optics has made remarkable strides in controlling and manipulating light through SPPs and other approaches[21,33,34], translating these capabilities to the X-ray regime presents significant challenges[35].

Efficient coupling between radiation and plasmons requires matching both energy and momentum. In the optical regime, energy conservation is achieved by matching the surface plasmon resonance frequency with the photon frequency. Momentum conservation demands matching the wavevector of the incident light with that of the SPP, a condition that can be achieved using prisms, gratings, or other optical structures that allow for precise wavevector engineering of the incident radiation[1,2,21,36]. However, coupling X-rays to SPPs is particularly challenging due to their extremely high frequency and wavevector, which far exceed the typical plasma frequency and SPP wavevectors. The absence of equivalent optical components to those in the optical range makes the direct matching between the wavevectors of X-rays and SPPs practically impossible. As a result, and despite the immense potential to advance fundamental science and enable novel applications, such as X-ray microscopy and spectroscopy, experimental demonstrations in this regime remain unrealized.

Here, we propose and demonstrate a novel approach to control the rate and emission angles of X-ray photons by exploiting the entanglement between X-rays and SPPs in the ultraviolet (UV) range. We leverage spontaneous parametric down-conversion (SPDC) of X-rays into longer-wavelength radiation[37–46] within an aluminum crystal to generate entangled photon pairs. One photon in each pair is an X-ray, while the other is a UV photon that can couple to surface plasmons in the crystal, effectively becoming an SPP. While the SPPs are absorbed within the crystal, the entangled X-ray photons emerge and can be detected. As we demonstrate, the rate and emission angle of these X-ray photons are influenced by the properties of the entangled SPPs. This unique relationship enables the imprinting of optical excitations onto the angular and energy spectrum of the X-ray signal, thereby allowing for precise control over their emission rate and direction.

Before proceeding, we note an important unique property of the effect described in this paper. The wavelength range of the long-wavelength photons/polaritons is 50-200 nm. For these wavelengths, the penetration depth of the polaritons or even the UV photons is shorter than the wavelength across almost the entire measured range. Therefore, surface effects are expected to be significant throughout the interaction length. In contrast, hard X-rays have wavelengths on the order of an Angstrom, resulting in an interaction volume that behaves effectively as bulk so that longitudinal phase matching must be taken into account. This is important because it enables the use of the reciprocal lattice vector, providing access to microscopic structural information [38,47,48].

In this work, we consider the conversion of X-rays into SPPs through SPDC. In general, SPDC involves a pump beam at frequency $\omega_p$ illuminating a nonlinear crystal to generate entangled photon pairs. Here, the input pump beam is in the X-ray range and leads to the generation of entangled X-ray and longer-wavelength photons. We refer to the X-ray photon as the signal and the longer-wavelength photon as the idler, with corresponding frequencies $\omega_s$ and $\omega_i$, respectively. When the frequency of the longer-wavelength photons approaches the surface plasmon resonance frequency, those photons with wavevectors matching the wavevector of the surface plasmons are strongly coupled to the surface plasmons, forming SPPs. Energy and momentum conservation of the SPDC process ensure strong angular correlation between the signal X-ray photon and the idler SPP, which are generated simultaneously.

As the SPP must adhere to a specific dispersion relation, determined by the properties of the materials forming the interface, the emission angles of the X-ray photons are consequently constrained by this dispersion relation. This process can be described using a macroscopic quantum electrodynamics (MQED) approach, which quantizes the macroscopic Maxwell's equations in a medium[49,50]. We employ this approach to calculate the count rate of the detected X-ray signal photons using:

$$\Gamma_s = \langle 0|\hat{a}_s^\dagger \hat{a}_s|0\rangle \qquad (1)$$

where $\hat{a}_s^\dagger$ and $\hat{a}_s$ are the creation and annihilation operators for the signal mode at the output of the crystal and are related to the corresponding electric field operators $\hat{E}_s^\dagger$ and $\hat{E}_s$ via the Poynting theorem. The $|0\rangle$ represents a vacuum state for both the signal and idler modes.

We begin by describing the theoretical model and writing the vectorial wave equation for the electric field operator, $\hat{E}_u(r,\omega_u)$, for a lossy, dispersive and inhomogeneous medium with a current source, $\hat{j}_u$, oscillating at a frequency $\omega_u$, where $u = s, i$ stands for signal / idler:

$$\nabla \times \nabla \times \hat{E}_u(\vec{r}, \omega_u) - \frac{\omega_u^2}{c^2}\varepsilon(\vec{r}, \omega_u)\hat{E}_u(\vec{r}, \omega_u) = -i\omega_u\mu_0\hat{j}_u(\vec{r}, \omega_u) \qquad (2)$$

where the position dependent permittivity, $\varepsilon(\vec{r}, \omega_u)$ is equal to unity (air) for $z < 0$ and the metal's complex permittivity, $\varepsilon_m(\omega_u)$, for $z > 0$ and we assume that the system is homogenous in the x-y plane. This equation accounts for linear and nonlinear interactions, as well as loss and quantum noise contributions, and applies to both signal and idler photons/polaritons. However, for the signal, we can employ the slowly varying envelope approximation, as shown in the supplementary information.

To model the down conversion of X-rays into long-wavelength radiation in absorbing media, we adopt a Langevin approach which introduces noise current operators, $\hat{j}_{N,u}$ to preserve the commutators. These noise operators are added to the nonlinear current operators, $\hat{j}_{NL,u}$, which couple the signal and idler electric field operators through a nonlinear coupling coefficient. The total source current in Eq. (2) is therefore: $\hat{j}_u = \hat{j}_{NL,u} + \hat{j}_{N,u}$ (see supplementary information).

The signal and idler electric fields are then found by solving Eq. (2). As the absorption length of the idler's electric field is comparable to or shorter than its wavelength, the contribution of the weak nonlinear interaction to the idler's propagation is negligible compared to that of the Langevin noise. In this case, the vacuum fluctuations of the idler electric field are proportional to the imaginary part of the Green's function, $G(r, r', \omega)$, that solves Eq. (2) (see supplementary information).

In the supplementary information we show that the solution of Eq. (2) for the signal electric field relies on the strongly absorbed idler electric field. It then follows that the signal's count rate, Eq. (1), is related to the above-mentioned Green's function via:

$$\Gamma_s = \int d\omega_i \int \frac{d^2 q_s}{(2\pi)^2} \frac{\hbar \omega_i^2}{\pi \epsilon_0 c^2} |C(q_s, q_p, \omega_i)|^2 \frac{q_i^2}{k_i^2(\omega_i)} \times$$
$$\int_L^0 dz' \int_L^0 dz\, e^{i\Delta k_{p,s}(z-z')} \text{Im}\{g_{q_i}(z, z', \omega_i)\} \quad (3)$$

where L is the interaction length for coherent SPDC, estimated from the width of the Bragg diffraction curve of the crystal. $k_i(\omega_i)$ is the idler wave vector in the crystal. $g_{q_i}$ is the Fourier transform of the mentioned Green's function $G(r, r', \omega_i) = \int \frac{d^2 \vec{q}_i}{(2\pi)^2} e^{i\vec{q}_i \cdot (\vec{\rho} - \vec{\rho}')} g_{q_i}(z, z', \omega_i)$, where $\vec{\rho}$ is the position along the crystal's surface and $\vec{q}_i$ is the idler wave-vector parallel to the crystal's surface. The Green's function $g_{q_i}(z, z', \omega_i)$ includes two contributions: one describing the field in an infinite homogeneous medium with the metal's permittivity, $\varepsilon_m(\omega_i)$, and the other describing the field reflected from the interface at $z = 0$.

The transverse wave vectors $\vec{q}_{j=p,s,i}$ (for pump, signal, and idler) satisfy transverse momentum conservation $\vec{q}_p = \vec{q}_s + \vec{q}_i$. The term $\Delta k_{p,s}$ describes a longitudinal mismatch along the z axis which includes only the pump and signal contributions. It is given by $\Delta k_{p,s} = k_{p,z}(q_p, \omega_{p,0}) + k_{s,z}(q_s, \omega_{p,0} - \omega_i) - G$ where $k_{p,z}$ and $k_{s,z}$ are the longitudinal pump and signal wave vectors, respectively, and $G$ is the reciprocal lattice vector used for phase matching. The contribution of the longitudinal idler wave vector, $e^{ik_{i,z}z}$, is accounted by the Green's function and thus included in the interaction's longitudinal phase matching also (see supplementary information). $C(q_s, q_p, \omega_i)$ is a pre-factor, which accounts for the propagation angles of the pump and signal photons and includes the nonlinear coupling coefficient of $\hat{j}_{NL,s}$.

Since the imaginary part of the Green's function is proportional to the vacuum fluctuations of the idler field, Eq. (3) indicates that the X-ray signal intensity depends directly on the vacuum fluctuations of the long-wavelength idler electric field. According to the fluctuation dissipation theorem, these fluctuations increase with the absorption of the idler field, reaching a maximum when $\hbar \omega_i \approx \hbar \omega_{SPR}$, where SPR stands for surface plasmon resonance. Consequently, from energy conservation, the signal count rate is expected to peak at an energy of $\hbar \omega_s = \hbar \omega_{pump} - \hbar \omega_{SPR}$, and the difference in the momentum of the pump and signal X-ray photons, $\vec{q}_p - \vec{q}_s$, follows that of a SPP.

The experimental setup is depicted in Fig. 1. We used a monochromatic pump beam to illuminate aluminum single-crystal. The pump photon energy was 9.978 keV for the results shown in Fig. 2 and 10.029 keV for those in Fig. 3. To facilitate the detection of weak SPDC signals emitted over a wide angular range, we employed a spherically bent crystal analyzer and a 2D pixelated x-ray detector. The aluminum crystal, analyzer, and detector were arranged in a Rowland circle geometry in the $\hat{x} - \hat{z}$ plane. This configuration provided both a large collection angle of about $6 \cdot 10^{-6}$ sr and a total energy resolution of 1.5 eV FWHM.

We chose aluminum as our sample material because of its relatively large SPR frequency, corresponding to an energy of approximately 10.65 eV. This choice enhances the signal-to-noise ratio (SNR) in our experiment, as we measure signal photons that are approximately 10.65 eV below the photon energy of the input beam. Given the energy resolution of our experimental setup, we could efficiently filter out noise from Bragg diffraction and Compton-scattered beams, effectively separating the SPR-enhanced peak from the elastic peak at energy $\hbar\omega_p$.

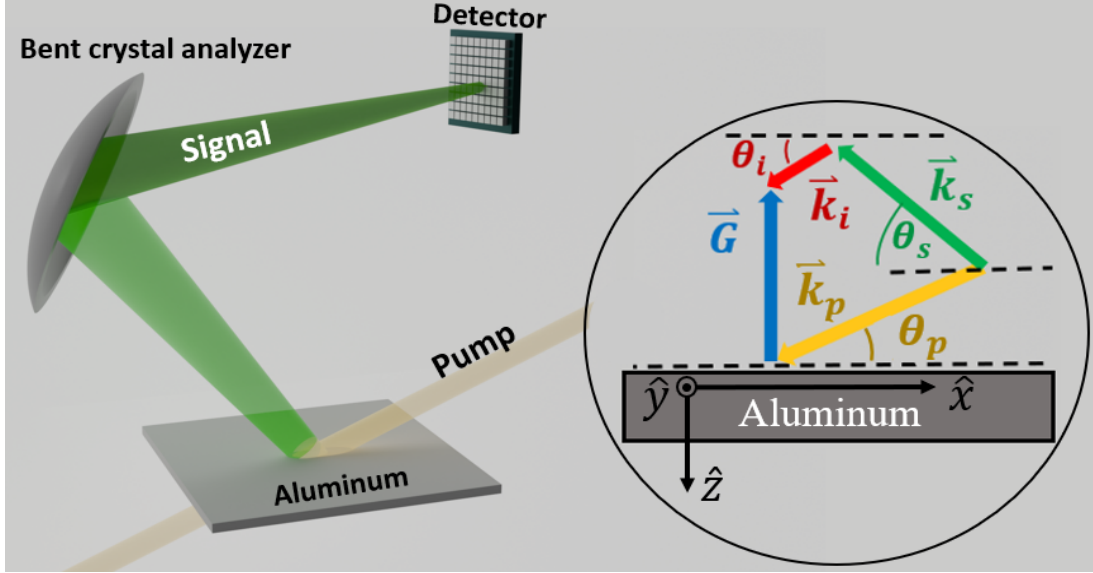

**Fig. 1| The experimental setup.** A monochromatic X-ray pump beam illuminates the aluminum crystal at an incidence angle $\theta_p$, generating signal photons and either idler photons or polaritons. The angles are measured relative to the atomic planes (dashed lines). A spherically bent crystal analyzer tuned to the signal frequency collects the X-ray signal photons and images them onto the detector. The inset shows a side view of the phase matching diagram with the idler (red) and the reciprocal lattice vector (blue) included.

To detect down-converted X-ray signal photons we used a near-Bragg geometry, where the signal photons emerge from the same surface of the crystal that the pump photons enter. This is also the surface along which the plasmon polaritons propagate. We denote $\theta_p$ and $\theta_s$ as the propagation angles of the pump and signal, respectively, with respect to the atomic planes as described in the inset in Fig. 1.

In the experiment, we measured the signal at the detector by varying both the crystal and detector arm angles. The crystal analyzer angle was tuned to select idler photon energies in the range of 6 to 25 eV for detection. This is achieved by tuning the analyzer to angles corresponding to lower photon energies of the signal, following the energy conservation relation $\hbar\omega_s = \hbar\omega_p - \hbar\omega_i$.

Figure 2(a) presents a colormap of the measured signal intensity as a function of idler energy and the deviation of the pump angle from the Bragg angle, obtained when using the reciprocal lattice vector normal to the (0,0,4) atomic planes for phase matching. Figure 2(b) shows the corresponding simulation results, calculated using Eq. 3.

The solid black and dashed white curves in Figs. 2(a) and 2(b) represent the kinematical solutions of the boundary condition, $\vec{q}_p = \vec{q}_s + \vec{q}_i$, for the pump angle (relative to the Bragg angle) when the idler dispersion, $q_i(\omega_i)$, follows that of an SPP propagating along an aluminum-air interface and a photon propagating in air, respectively. The dashed-white line, referred to as the dielectric light line[1,17,51], highlights the influence SPPs exert on the angular spectrum of the X-ray beam.

As observed in Fig. 2(a), the measured pump angle of maximum intensity transitions from the right to the left of the dielectric light line as the idler energy crosses the SPR energy, $\hbar\omega_{SPR}$. This transition results in two distinct intensity branches, which agree well with our theoretical predictions in Fig. 2(b). The dielectric light line asymptotically approaches the two branches, separating optically bound modes (with momentum greater than that in vacuum) from radiative modes (with momentum smaller than or equal to that in vacuum)[1,17,51]. **The observed angular dependence of the pump beam is a key signature of plasmonic systems and has been observed in other types of measurements**[1,17,51]. In our work, this phenomenon arises from phase matching for the SPDC, where, near the SPR resonance, it is governed by the SPP dispersion, while away from resonance, it is determined by normal dispersion.

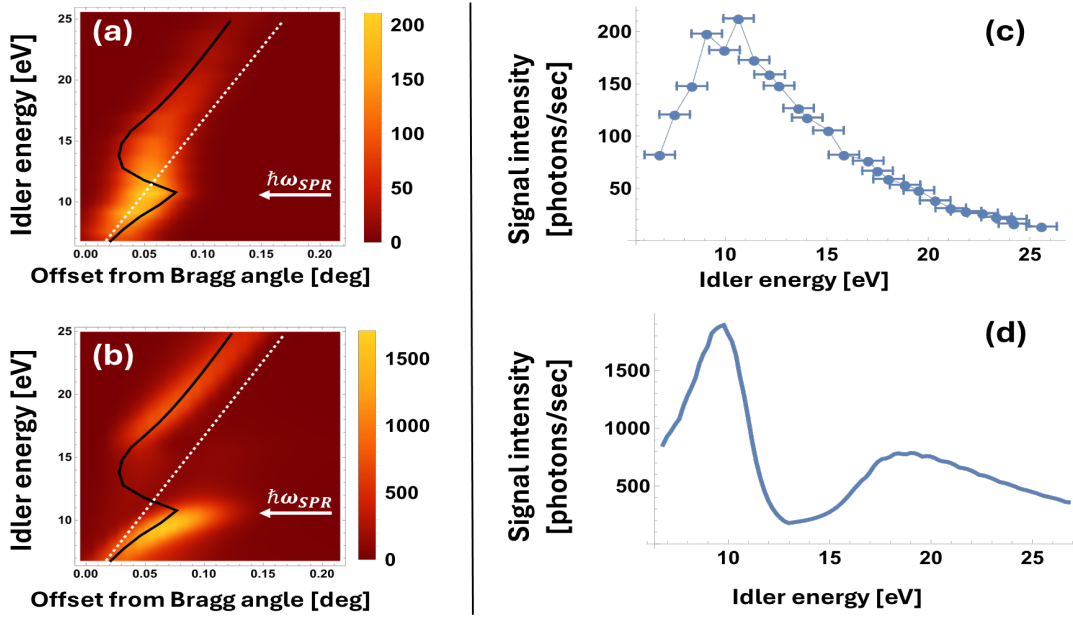

**Fig. 2| SPDC of X-rays into UV plasmon polaritons in Aluminum, obtained using a 9.978 KeV pump photons and the reciprocal lattice vector normal to the (0,0,4) atomics planes. (a)** Experimental results and **(b)** QED simulation of the signal intensity as a function of the idler photon energy and the deviation of the pump angle from the Bragg angle. The solid-black and dashed-white curves represent kinematical solutions to the boundary condition, $\vec{q}_p = \vec{q}_s + \vec{q}_i$, for the pump angle (relative to the Bragg angle), where the idler's dispersion, $q_i(\omega_i)$, follows that of a SPP and a photon propagating in vacuum, respectively. **(c)** and **(d)** show the signal intensity as a function of the idler photon energy. **(c)** Dots represent experimental results and **(d)** line simulation. **(c)** The horizontal error bars represent the experimental energy uncertainty, which corresponds to the total energy resolution. The vertical error is smaller than the point size.

Next, we plot the signal spectrum by integrating the intensity in Fig. 2(a) and Fig. 2(b) over the pump angle for each idler energy. Both theory and experiment indicate that the highest signal intensity occurs at the SPR frequency and gradually decreases for optically radiative modes, i.e., for idler frequencies higher than the bulk plasmon resonance frequency, $\hbar\omega_{BPR} \approx 15.8\ eV$. **The observed angular and energy dependence of the X-ray photons provides clear experimental evidence for the existence of SPPs and their coupling to the X-ray photons**.

Following the observation of SPDC of X-ray photons into SPPs for the (0,0,4) atomic planes, we verified that the SPR enhancement is not specific to a particular atomic plane or pump energy by measuring the signal for the Al(1,1,1) and Al(0,0,2) reflections at a pump photon energy of 10.029 KeV, as shown in Fig. 3. To further confirm that the enhancement is independent of detector alignment, we conducted measurements at different detector angles. The pump was tuned away from the Bragg angle by 0.19°. The experimental spectra for Al(1,1,1) and Al(0,0,2) are shown in Figs. 3(a) and 3(b), respectively, while the corresponding simulations based on Eq. 3 are presented in Figs. 3(c) and 3(d), respectively.

As shown in the experimental spectra in Figs. 3(a) and 3(b), an enhancement is observed at a signal energy which is lower than the new pump energy (of 10.029 KeV) by about $\hbar\omega_{SPR} = 10.65\ eV$, in accordance with the observation for the Al(0,0,4) atomic planes with a pump energy of 9978 eV, as demonstrated in Fig. 2.

In the theoretical spectra, Figs. 3(c) and 3(d), a second pronounced peak is observed at the radiative idler energies (i.e., idler energies larger than $\hbar\omega_{BPR}$). A corresponding, albeit less pronounced, signal is also observed in the experimental data, Figs. 3(a) and 3(b). This feature was consistently observed across multiple experimental runs.

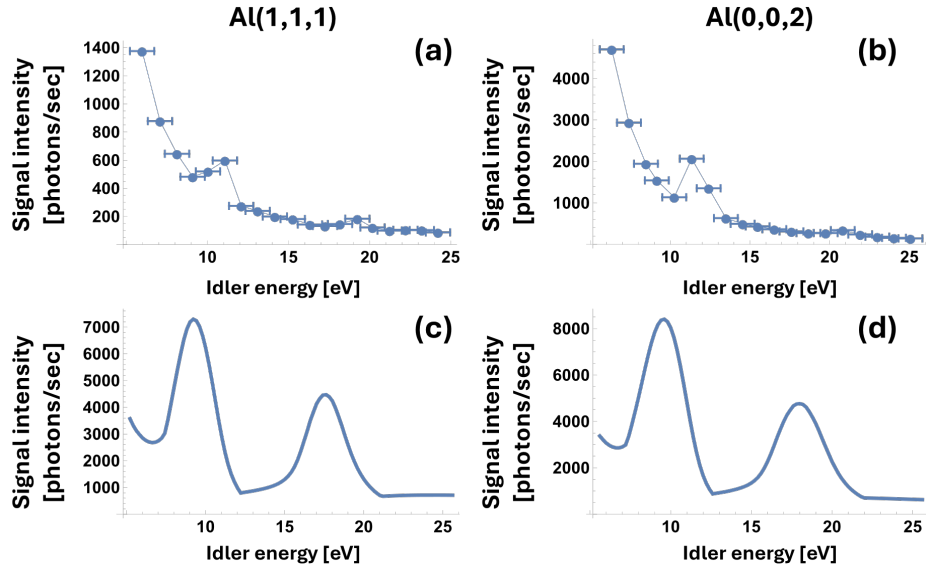

**Fig. 3| Experimental (a–b) and simulated (c-d) signal spectra for the Al(1,1,1) and Al(0,0,2) atomic planes, obtained using a 10.029 KeV pump photons. (a)** and **(b)** The horizontal error bars represent the experimental energy uncertainty, which corresponds to the total energy resolution. The vertical error is smaller than the point size. See the text for additional details.

The similarity between the experimental spectrum of the Al(1,1,1) and Al(0,0,2) reflections, Figs. 3(a) and 3(b), arises due to the common pump angle deviation and the similarity of the reciprocal lattice vector used for those measurements. This is also evident from their corresponding theoretical spectra, Figs. 3(c) and 3(d).

According to our theoretical model, a resonance enhancement is expected near the bulk plasmon resonance energy arising from the divergence of $\frac{1}{k_i^2(\omega_i)}$ when $\omega_i = \omega_{BPR}$. But, like the SPR, the exact idler energy for this bulk resonance is not exactly at $\hbar\omega_{BPR}$. This is due to the influence of the SPDC phase-mismatch function on the signal's spectrum.

It is expected that this resonance enhancement in the radiative idler energies will appear in both the sample scan (Fig. 2) and the detector scan spectra (Fig. 3). However, analyzing the SPDC phase-mismatch function reveals that the portion of signal angles satisfying $\Delta k_z = 0$ for idler energies just above $\hbar\omega_{BPR}$ is larger in the detector scans compared to the sample scan. This explains the differences between the radiative spectral shape (idler energies above $\hbar\omega_{BPR}$) observed in the sample scan (Fig. 2(d)) and in the detector scans (Figs. 3(c) and 3(d)).

Overall, both Fig. 2 and Fig. 3 show good qualitative agreement between the experimental results and simulations. Several experimental factors contribute to the small deviations observed between the experiment and theory. These include the photon energy bandwidth of the input beam, aperture sizes, input flux, and the lack of precise surface control for the crystal. The mosaicity of the crystal is an important factor, which we could only estimate. A high degree of mosaicity shortens the coherence length, which significantly affects the bulk contribution to the SPDC signal. This is because the bulk contribution to the nonlinearity is more dependent on the interaction length compared to the surface contribution[17]. The scattered Bragg curve widths, which were about 0.17°-0.21° for the reflections in our experiment, suggest low crystalline quality material with a coherent SPDC interaction length no larger than 60 nm. The short coherence length of the aluminum resulted in a greater contribution from surface nonlinearity.

In conclusion, we have presented a method for controlling X-ray flux and emission angles through entanglement with longer-wavelength photons, which efficiently interact with surface plasmons. Although the entanglement cannot be directly measured—since the SPP cannot emerge from the crystal—the strong correlation between the SPP and the entangled X-ray photon persists. As a result, the rate and emission angle of the X-rays are controlled by the properties of the SPP.

Enhanced surface control would allow greater precision in regulating the emitted signal count rate and emission angles. This can be achieved through nanofabrication techniques commonly used in metamaterials and nano-optics[16,51]. Furthermore, as X-rays are highly sensitive to local fields with atomic-scale resolution, our method opens new frontiers for studying the effects of nanofabrication on local fields in a noninvasive manner. This information, currently inaccessible, could significantly enhance the understanding of nanostructure functionality and offer valuable insights into their optimization.

# Methods

**Experimental setup**

The experiment was conducted at the GALAXIES beamline at the SOLEIL synchrotron facility on the RIXS endstatation[52,53]. The source was a U20 (20mm) period undulator, 98 periods. The energy of the input pump beam was selected by the Si(111) double-crystal monochromator of the beamline and the signal energy was selected using a 1m radius Si(555) spherically bent crystal analyzer. The total photon energy resolution of the systems was 1.5 ±0.1 eV full width at half maximum (FWHM). The pump beam was focused to a $30 \mu m (\hat{x}$ direction$) \times 90 \mu m (\hat{y}$ direction$)$ spot onto the aluminum sample. The input beam intensity was approximately $1.5 \times 10^{12} ph/sec$. The detector used was a hybrid pixelated 2D detector with 55 μm pixel size. A helium flight path was inserted between the sample, analyzer and detector to eliminate air absorption of the x-rays. In addition, an adjustable 5 mm iris was placed in front of the analyzer to block the diffracted beam (and its tail) from entering the detector.

**Experimental and analysis procedure**

*Experimental technique and analysis for Figs. 2 (a) and 2(c)*

First, we tuned the sample and detector to the Bragg angles, then detuned the detector arm by 0.2° from the Bragg condition. The crystal analyzer angle was adjusted to select a range of idler photon energies, corresponding to lower signal photon energies as dictated by the energy conservation relation $\hbar\omega_s = \hbar\omega_p - \hbar\omega_i$. For each idler photon energy, we scanned the sample angle relative to the pump beam (source), and recorded the signal intensity using the detector.

To separate the SPDC signal from the uniformly distributed background noise, we subtracted the background signal, measured in a different region of interest on the detector, from the signal at the analyzer's focal point. The SPDC signal was then extracted by removing the noise component from the total recorded signal. By repeating this process for the whole idler energy range, we reconstructed the experimental intensity map shown in Fig. 2(a). The spectrum presented in Fig. 2(c) was obtained by taking the maximum intensity at each idler energy.

*Experimental technique and analysis for Figs. 3 (a-c)*

The procedure for obtaining the results presented in Fig. 3 was the same as described above, except that the pump angle was offset by 0.19° and the detector arm was scanned over a specified range of 0.4°.

# References

1. Maier, S. A. Plasmonics: Fundamentals and applications. *Plasmonics: Fundamentals and Applications* 1–223 (2007).

2. Raether, H., Hohler, G. & Niekisch, E. A. Surface Plasmons on Smooth and Rough Surfaces and on Gratings. *Springer Tracts in Modern Physics* **111**, 136 (1988).

3. Ritchie, R. H. Plasma Losses by Fast Electrons in Thin Films. *Physical Review* **106**, 874–881 (1957).

4. Garcia-Vidal, F. J. *et al.* Spoof surface plasmon photonics. *Rev Mod Phys* **94**, (2022).

5. Yu, H., Peng, Y., Yang, Y. & Li, Z. Y. Plasmon-enhanced light–matter interactions and applications. *npj Computational Materials 2019 5:1* **5**, 1–14 (2019).

6. Stockman, M. I. *et al.* Roadmap on plasmonics. *Journal of Optics* **20**, 043001 (2018).

7. Moreland, J., Adams, A. & Hansma, P. K. Efficiency of light emission from surface plasmons. *Phys Rev B* **25**, 2297–2300 (1982).

8. Kneipp, K. *et al.* Single Molecule Detection Using Surface-Enhanced Raman Scattering (SERS). *Phys Rev Lett* **78**, 1667–1670 (1997).

9. Okamoto, K. *et al.* Surface-plasmon-enhanced light emitters based on InGaN quantum wells. *Nature Materials 2004 3:9* **3**, 601–605 (2004).

10. Sun, G., Khurgin, J. B. & Soref, R. A. Practicable enhancement of spontaneous emission using surface plasmons. *Appl Phys Lett* **90**, (2007).

11. Fort, E. & Grésillon, S. Surface enhanced fluorescence. *J Phys D Appl Phys* **41**, 013001 (2007).

12. Ivanov, K. A. *et al.* Control of the surface plasmon dispersion and Purcell effect at the metamaterial-dielectric interface. *Scientific Reports 2020 10:1* **10**, 1–8 (2020).

13. Reinisch, R. & Nevière, M. Electromagnetic theory of diffraction in nonlinear optics and surface-enhanced nonlinear optical effects. *Phys Rev B* **28**, 1870–1885 (1983).

14. Chen, C. K., De Castro, A. R. B. & Shen, Y. R. Surface-Enhanced Second-Harmonic Generation. *Phys Rev Lett* **46**, 145–148 (1981).

15. Quail, J. C., Rako, J. G., Simon, H. J. & Deck, R. T. Optical Second-Harmonic Generation with Long-Range Surface Plasmons. *Phys Rev Lett* **50**, 1987–1989 (1983).

16. Kauranen, M. & Zayats, A. V. Nonlinear plasmonics. *Nature Photonics 2012 6:11* **6**, 737–748 (2012).

17. Y.R. Shen. Principles of nonlinear optics. *Principles of nonlinear optics* 1–39 (1873).

18. Ni, X., Emani, N. K., Kildishev, A. V., Boltasseva, A. & Shalaev, V. M. Broadband light bending with plasmonic nanoantennas. *Science (1979)* **335**, 427 (2012).

19. Hong, M., Dawkins, R. B., Bertoni, B., You, C. & Magaña-Loaiza, O. S. Nonclassical near-field dynamics of surface plasmons. *Nature Physics 2024 20:5* **20**, 830–835 (2024).

20. Dolev, I., Epstein, I. & Arie, A. Surface-plasmon holographic beam shaping. *Phys Rev Lett* **109**, (2012).

21. Barnes, W. L., Dereux, A. & Ebbesen, T. W. Surface plasmon subwavelength optics. *Nature 2003 424:6950* **424**, 824–830 (2003).

22. Pendry, J. B., Martín-Moreno, L. & Garcia-Vidal, F. J. Mimicking Surface Plasmons with Structured Surfaces. *Science (1979)* **305**, 847–848 (2004).

23. Kildishev, A. V., Klar, T. A., Drachev, V. P. & Shalaev, V. M. Thin metal-dielectric nanocomposites with a negative index of refraction. *Nanophotonics with Surface Plasmons* 271–308 (2007).

24. Kumar, G. & Sarswat, P. K. Interaction of Surface Plasmon Polaritons with Nanomaterials. 103–129 (2016).

25. Schuller, J. A. *et al.* Plasmonics for extreme light concentration and manipulation. *Nature Materials 2010 9:3* **9**, 193–204 (2010).

26. Jain, P. K., Huang, X., El-Sayed, I. H. & El-Sayed, M. A. Noble metals on the nanoscale: Optical and photothermal properties and some applications in imaging, sensing, biology, and medicine. *Acc Chem Res* **41**, 1578–1586 (2008).

27. Homola, J., Yee, S. S. & Gauglitz, G. Surface plasmon resonance sensors: review. *Sens Actuators B Chem* **54**, 3–15 (1999).

28. Zayats, A. V. & Smolyaninov, I. I. Near-field photonics: surface plasmon polaritons and localized surfaceplasmons. *Journal of Optics A: Pure and Applied Optics* **5**, S16 (2003).

29. Chen, C. K., De Castro, A. R. B., Shen, Y. R. & DeMartini, F. Surface Coherent Anti-Stokes Raman Spectroscopy. *Phys Rev Lett* **43**, 946–949 (1979).

30. Nie, S. & Emory, S. R. Probing Single Molecules and Single Nanoparticles by Surface-Enhanced Raman Scattering. *Science (1979)* **275**, 1102–1106 (1997).

31. Willets, K. A. & Van Duyne, R. P. Localized surface plasmon resonance spectroscopy and sensing. *Annu Rev Phys Chem* **58**, 267–297 (2007).

32. Wang, X., Huang, S. C., Hu, S., Yan, S. & Ren, B. Fundamental understanding and applications of plasmon-enhanced Raman spectroscopy. *Nature Reviews Physics 2020 2:5* **2**, 253–271 (2020).

33. Maier, S. A. Plasmonics: Fundamentals and applications. *Plasmonics: Fundamentals and Applications* 1–223 (2007).

34. Gramotnev, D. K. & Bozhevolnyi, S. I. Plasmonics beyond the diffraction limit. *Nature Photonics 2010 4:2* **4**, 83–91 (2010)

35. Röhlsberger, R., Evers, J. & Shwartz, S. Quantum and Nonlinear Optics with Hard X-Rays. *Synchrotron Light Sources and Free-Electron Lasers: Accelerator Physics, Instrumentation and Science Applications* 1399–1431 (2020)

36. Kretschmann, E. & Raether, H. Radiative Decay of Non Radiative Surface Plasmons Excited by Light. *Zeitschrift fur Naturforschung - Section A Journal of Physical Sciences* **23**, 2135–2136 (1968).

37. Danino, H. & Freund, I. Parametric Down Conversion of X Rays into the Extreme Ultraviolet. *Phys Rev Lett* **46**, 1127–1130 (1981).

38. Tamasaku, K., Sawada, K., Nishibori, E. & Ishikawa, T. Visualizing the local optical response to extreme-ultraviolet radiation with a resolution of λ/380. *Nature Physics 2011 7:9* **7**, 705–708 (2011).

39. Borodin, D., Levy, S. & Shwartz, S. High energy-resolution measurements of x-ray into ultraviolet parametric down-conversion with an x-ray tube source. *Appl Phys Lett* **110**, (2017).

40. S. Sofer, O. Sefi, E. Strizhevsky, H. Aknin, S. P. Collins, G. Nisbet, B. Detlefs, Ch. J. Sahle & S. Shwartz. Observation of strong nonlinear interactions in parametric down-conversion of X-rays into ultraviolet radiation. *Nature Communications 2019 10:1* **10**, 1–8 (2019).

41. A. Schori, C. Bömer, D. Borodin, S. P. Collins, B. Detlefs, M. Moretti Sala, S. Yudovich, and S. Shwartz. Parametric Down-Conversion of X Rays into the Optical Regime. *Phys Rev Lett* **119**, (2017).

42. Borodin, D., Schori, A., Rueff, J. P., Ablett, J. M. & Shwartz, S. Evidence for Collective Nonlinear Interactions in X Ray into Ultraviolet Parametric Down-Conversion. *Phys Rev Lett* **122**, (2019).

43. Cohen, R. & Shwartz, S. Theory of nonlinear interactions between x rays and optical radiation in crystals. *Phys Rev Res* **1**, (2019).

44. Tamasaku, K. & Ishikawa, T. Interference between compton scattering and X-ray parametric down-conversion. *Phys Rev Lett* **98**, (2007).

45. Tamasaku, K., Sawada, K. & Ishikawa, T. Determining X-Ray nonlinear susceptibility of diamond by the optical fano effect. *Phys Rev Lett* **103**, (2009).

46. Barbiellini, B., Joly, Y. & Tamasaku, K. Explaining the x-ray nonlinear susceptibility of diamond and silicon near absorption edges. *Phys Rev B Condens Matter Mater Phys* **92**, (2015).

47. Glover, T. E. *et al.* X-ray and optical wave mixing. *Nature 2012 488:7413* **488**, 603–608 (2012).

48. Freund, I. & Levine, B. F. Optically Modulated X-Ray Diffraction. *Phys Rev Lett* **25**, 1241–1245 (1970).

49. Rivera, N. & Kaminer, I. Light–matter interactions with photonic quasiparticles. *Nature Reviews Physics 2020 2:10* **2**, 538–561 (2020).

50. Scheel, S. & Buhmann, S. Y. Macroscopic quantum electrodynamics-concepts and applications. *Acta Physica Slovaca* **58**, 675–809 (2008).

51. Novotny, L. & Hecht, B. Principles of Nano-Optics. *Principles of Nano-Optics* **9781107005464**, 1–564 (2012).

52. Ablett, J. M. *et al.* The GALAXIES inelastic hard X-ray scattering end-station at Synchrotron SOLEIL. *urn:issn:1600-5775* **26**, 263–271 (2019).

53. Rueff, J. P. *et al.* The GALAXIES beamline at the SOLEIL synchrotron: inelastic X-ray scattering and photoelectron spectroscopy in the hard X-ray range. *urn:issn:1600-5775* **22**, 175–179 (2015).

# Surface Plasmon-Enhanced X-ray Ultraviolet Nonlinear Interactions

## Supplementary information

*H. Aknin[1], O. Sefi[1], D. Borodin[1], J.-P. Rueff[2, 3], J. Ablett[2], and S. Shwartz[1]\**

[1]*Physics Department and Institute of Nanotechnology, Bar-Ilan University, Ramat Gan, 52900 Israel*

[2] *Synchrotron SOLEIL, L'Orme des Merisiers, Départementale 128, 91190 Saint-Aubin, France*

[3] *Sorbonne Université, CNRS, Laboratoire de Chimie Physique-Matière et Rayonnement, LCPMR, F-75005 Paris, France*

## Calculation of Signal Photon Detection Rate

In this section, we outline the derivation of Eq. (3) of the main text, which we used to compute the count rate at the signal frequency, and provide the essential details required for the numerical calculations.

### A. The classical wave equations for the idler and signal electric fields

We begin by examining the classical wave equations that describe the electric fields of the signal and idler. We assume that the interaction is weak and apply the undepleted pump approximation. Under this approximation, the pump is absorbed in the sample, but we neglect the energy transfer from it to the signal and idler beams.

To describe the wave dynamics of the idler field at a scale smaller than its wavelength, the idler electric field must obey the following vectorial wave equation (Eq. (2) in the main text):

$$\nabla \times \nabla \times \vec{E}_i(r,\omega_i) - \frac{\omega_i^2}{c^2}\varepsilon(r,\omega_i)\vec{E}_i(r,\omega_i) = -i\omega_i\mu_0\vec{j}_i(r,\omega_i). \quad (s1)$$

Here the space dependent macroscopic permittivity is given by $\varepsilon(r,\omega_i) = \begin{cases} 1 & z < 0 \\ \varepsilon_m(\omega_i) & z > 0 \end{cases}$, where $\varepsilon_m(\omega_i)$ is the metal permittivity, and the sample is assumed to be in air. $\vec{j}_i(r,\omega_i)$ represents the current density acting as the source term.

Using the following Fourier Transforms:

$$\vec{E}(r,\omega) = \int \frac{d^2q}{(2\pi)^2} e^{i\vec{q}\cdot\vec{\rho}} \vec{E}(q,\omega,z) \quad (s2.1)$$

$$\vec{j}(r,\omega) = \int \frac{d^2q}{(2\pi)^2} e^{i\vec{q}\cdot\vec{\rho}} \vec{j}(q,\omega,z), \quad (s2.2)$$

where $\vec{q} = (k_x, k_y)$ is the wave vector and $\vec{\rho} = (x, y)$ is position in the transverse direction, both parallel to the surface of the crystal (See Fig. 1 in the main text), we obtain:

$$(\partial_z \hat{z} + i\vec{q}_i) \times (\partial_z \hat{z} + i\vec{q}_i) \times \vec{E}_i(q_i, \omega_i, z) - \frac{\omega_i^2}{c^2} \varepsilon_m(\omega_i) \vec{E}_i(q_i, \omega_i, z) \quad (s3)$$
$$= -i\omega_i \mu_0 \vec{j}_i(q_i, \omega_i, z).$$

In contrast to the idler field, the absorption length of X-rays is much greater than their wavelength and their refractive index is close to unity. In this case, the propagation of the signal field can be described by a scalar wave equation. Applying the slowly varying envelope approximation (SVEA), the equation for the signal's electric field reads:

$$\partial_z \vec{E}_s(q_s, \omega_s, z) = \omega_s \mu_0 \frac{\vec{j}_s(q_s, \omega_s, z)}{2 k_{s,z}(q_s, \omega_s)} e^{i k_{s,z}(q_s, \omega_s) z}, \quad (s4)$$

where $k_{s,z} = \sqrt{\varepsilon_m(\omega_s) \frac{\omega_s^2}{c^2} - q_s^2}$.

### B. Nonlinear current density

The source term, $\vec{j}_{u=i,s}$ in Eq. (s3) and Eq. (s4) includes the nonlinear current density resulting from the interaction between the X-rays and the sample. The sample can be described as a dense, cold plasma, with electrons exhibiting the periodicity of the crystal[1,2].

Assuming a monochromatic, plane wave, pump electric field oscillating at frequency $\omega_{p,0}$ the nonlinear current density in the direction of the $u^{th}$ field is given by[3,4]:

$$j_u^{(2)}(\vec{q}_u, \omega_u, z) = \quad (s5)$$
$$\int d\omega_v \, \delta(\omega_{p,0} - \omega_v - \omega_u) \int d^2 q_v \, \delta(\vec{q}_u + \vec{q}_v - \vec{q}_p) \times$$
$$\times \sigma(\omega_s, \omega_i) \omega_u \vec{E}_v^*(\vec{q}_v, \omega_v, z) |\vec{E}_{p,0}| e^{iz(k_{p,z} - G)} (\hat{e}_i^* \cdot \vec{G})(\hat{e}_p \cdot \hat{e}_s),$$

where the subindices are either $u = s$ and $v = i$, or $u = i$ and $v = s$. The delta function $\delta(\omega_{p,0} - \omega_v - \omega_u)$ arises from energy conservation requirement while $\delta(\vec{q}_u + \vec{q}_v - \vec{q}_p)$ reflects the continuity of the electric fields along the transverse dimension (the x-y plane). $|\vec{E}_{p,0}|$ is the amplitude of the pump electric field, $\vec{G}$ is the reciprocal lattice vector and we assume it is oriented along the negative z direction (symmetric Bragg geometry). $\hat{e}_{l=p,s,i}$ represents a unit vector in the direction of the pump/ signal/ idler electric field, respectively. $\sigma = \frac{e^2 \rho_G}{m^2 \omega_{p,0} \omega_i^2 \omega_s} \ll 1$ is the nonlinear coefficient with $\rho_G$ the G-th Fourier component of the unperturbed charge density, $\rho_0$, given by: $\rho_G = \frac{1}{V} \int_V \rho_0 \, e^{-i\vec{G}\cdot\vec{r}} d^3 r$, where the integration is performed over the volume, $V$, of the unit cell[3].

### C. Quantization of the Electric Field and Source Terms

After deriving the classical forms of the formalism, we turn to the quantum formulation of the problem, as SPDC is initiated by the vacuum fluctuations of the electromagnetic field.

The major difference between the classical and quantum formalisms is that, in the classical formalism, the vectors commute, whereas in the quantum formalism, the operators must obey the bosonic commutation relations.

We start with the signal.

We define the commutators of the bosonic signal electric field propagating at an angle $\theta_s$ with respect to the atomic planes (see Fig. 1 in the main text) as:

$$[\hat{E}_s(q_s', z', \omega_s'), \hat{E}_s^\dagger(q_s, z, \omega_s)] = \frac{2\hbar\omega_s \eta(\omega_s)}{\sin(\theta_s)} \delta(q_s - q_s')\delta(z - z')\delta(\omega_s - \omega_s') \quad (s6.1)$$

$$[\hat{E}_s(q_s', z', \omega_s'), \hat{E}_s(q_s, z, \omega_s)] = 0 \quad (s6.2)$$

The corresponding creation and annihilation operators are $\hat{a}_s^\dagger$ and $\hat{a}_s$, respectively, and they satisfy the commutators:

$$[\hat{a}_s(q_s', z', \omega_s'), \hat{a}_s^\dagger(q_s, z, \omega_s)] = \frac{1}{(2\pi)^3} \delta(q_s - q_s')\delta(z - z')\delta(\omega_s - \omega_s') \quad (s7.1)$$

$$[\hat{a}_s(q_s', z', \omega_s'), \hat{a}_s(q_s, z, \omega_s)] = 0. \quad (s7.2)$$

The signal electric field operator is then related to the annihilation operator as follows:

$$\hat{E}_s(q_s, \omega_s, z) = \beta(q_s, \omega_s)\hat{a}_s(q_s, \omega_s, z), \quad (s8)$$

where we denote $\beta(q_s, \omega_s) \equiv \sqrt{\frac{2\hbar\omega_s \eta(\omega_s)}{\sin(\theta_s)}}$ and $\sin(\theta_s) \equiv \frac{\sqrt{\varepsilon_m(\omega_s)\frac{\omega_s^2}{c^2} - q_s^2}}{\sqrt{\varepsilon_m(\omega_s)}\frac{\omega_s}{c}}$. The pre-factor $\beta$ is introduced to ensure Poyinting theorem is satisfied[5].

In the presence of loss in the system, we need to introduce a Langevin bosonic noise operator, $\hat{f}$. This ensures that the commutators of the electric fields and their corresponding bosonic operators remain invariant during propagation. They satisfy the bosonic commutation relations:

$$[\hat{f}_j(q_s', z', \omega_s'), \hat{f}_k^\dagger(q_s, z, \omega_s)] = \frac{1}{(2\pi)^3} \delta_{j,k}\delta(q_s - q_s')\delta(z - z')\delta(\omega_s - \omega_s'), \quad (s9.1)$$

and

$$[\hat{f}_j(q_s', z', \omega_s'), \hat{f}_k(q_s, z, \omega_s)] = 0. \quad (s9.2)$$

Since this operator represents a noise term, it can be incorporated into the wave equation by treating it as a current source term, as shown in[6,7]:

$$\hat{j}_N(q, \omega, z) = N(\omega)\hat{f}(q, \omega, z) \quad (s10)$$

where $N(\omega) = \sqrt{\epsilon_0 4\pi\hbar\omega^2 \varepsilon_{m,I}(\omega)}$ is introduced to ensure the correlation function of $\hat{j}_N$, $\langle 0|\hat{j}_N(q,\omega,z)\hat{j}_N^\dagger(q,\omega,z)|0\rangle$ agrees with the fluctuation–dissipation theorem for the medium.

Using Eq. (s5) for the nonlinear current and Eq. (s10) for the noise current we obtain a Heisenberg Langevin equation for the signal:

$$\partial_z \hat{E}_s(q_s,\omega_s,z) = \omega_s \mu_0 \frac{\sigma(\omega_s,\omega_i)\omega_s \hat{E}_i^\dagger(\vec{q}_i,\omega_i,z)|\vec{E}_{p,0}|e^{iz(k_{p,z}-G)}(\hat{e}_i^* \cdot \vec{G})(\hat{e}_p \cdot \hat{e}_s)}{2k_{s,z}(q_s,\omega_s)} e^{ik_{s,z}(q_s,\omega_s)z} + \\ +\omega_s \mu_0 \frac{\sqrt{\epsilon_0 4\pi\hbar\omega_s^2 \varepsilon_{m,I}(\omega_s)}}{2k_{s,z}(q_s,\omega_s)} \hat{f}_s(q_s,\omega_s,z)e^{ik_{s,z}(q_s,\omega_s)z},$$  (s11)

where we used the conservation relations: $\omega_i = \omega_{p,0} - \omega_s$ and $\vec{q}_i = \vec{q}_p - \vec{q}_s$.

Moving on to quantize the wave equation of the idler's electric field, we define the commutation relations for the idler's electric field as follows[6,7]:

$$\left[\hat{E}_{i,\alpha'}(q_i',z',\omega_i'), \hat{E}_{i,\alpha}^\dagger(q_i,z,\omega_i)\right] = \frac{\hbar\omega_i^2}{\pi\epsilon_0 c^2} \operatorname{Im} g_{q_i}^{(\alpha',\alpha)}(z',z,\omega_i)\,\delta(q_i - q_i')\delta(\omega_i - \omega_i') \quad (s12.1)$$

$$\left[\hat{E}_{i,\alpha'}(q_i',z',\omega_i'), \hat{E}_{i,\alpha}(q_i,z,\omega_i)\right] = 0. \quad (s12.2)$$

Eq. (s12) is consistent with the fluctuation-dissipation relation for the idler electric field in the medium, where dissipation is characterized by the imaginary part of the Green's tensor component $g_{q_i}^{(\alpha',\alpha)}$. The tensor $g_{q_i}^{(\alpha',\alpha)}$ is the solution of

$$(\partial_z \hat{z} + i\vec{q}_i) \times (\partial_z \hat{z} + i\vec{q}_i) \times g_{q_i}(z,z',\omega_i) - \frac{\omega_i^2}{c^2}\varepsilon_m(\omega_i)g_{q_i}(z,z',\omega_i) = \bar{I}\delta(z-z'), \quad (s13)$$

with the boundary conditions for the idler electric field at the metal-air interface (at $z = 0$) and at infinity[8,9].

Next, we note that the analysis for the idler electric field can be significantly simplified, given that at the idler photon energies investigated in this study, the absorption length of the idler is comparable to or shorter than its wavelength. In this case, the contribution of the weak nonlinear interaction (with coupling $\sigma \ll 1$) to the idler's propagation is negligible compared to that of Langevin noise. In this case of strong absorption, the wave equation for idler electric field operator, $\hat{E}_i$ can be written as:

$$(\partial_z \hat{z} + i\vec{q}_i) \times (\partial_z \hat{z} + i\vec{q}_i) \times \hat{E}_i(q_i,\omega_i,z) - \frac{\omega_i^2}{c^2}\varepsilon_m(\omega_i)\hat{E}_i(q_i,\omega_i,z) \\ = -i\omega_i\mu_0 \hat{j}_N(q_i,\omega_i,z). \quad (s14)$$

By substituting Eq. (s10) for $\hat{j}_{N,i}$ and applying Green's function, the idler electric field operator, $\hat{E}_i$, can be expressed as:

$$\hat{E}_{i,\alpha}(q_i,\omega_i,z) = \int dz' \, (i\mu_0\omega_i)\sqrt{\epsilon_0 4\pi\hbar\omega_i^2 \varepsilon_{m,I}(\omega_i)} g_{q_i}^{(\alpha,\beta)}(z,z',\omega_i)\hat{f}_{i,\beta}(q_i,\omega_i,z). \quad (s15)$$

Importantly, and in accordance with the commutation relations for $\hat{E}_i$, Eq. (s15) indicates that the *vacuum fluctuations* of the idler electric fields are related to the Green's function via the fluctuation dissipation relation[6,7]:

$$\langle vac|\hat{E}_{i,\alpha'}(q_i', z', \omega_i')\hat{E}_{i,\alpha}^\dagger(q_i, z, \omega_i)|vac\rangle$$
$$= \frac{\hbar\omega_i^2}{\pi\epsilon_0 c^2}\, \text{Im}\, g_{q_i}^{(\alpha',\alpha)}(z', z, \omega_i)\, \delta(q_i - q_i')\delta(\omega_i - \omega_i'). \quad (s16)$$

### D. Calculation of the signal operators at the output of the sample

We are interested in the count rate of the signal X-ray photons. For simplicity, we neglect propagation effects between the sample's output and the detector.

Substituting the relation between $\hat{E}_s$ and $\hat{a}_s$, Eq. (s8), into the Heisenberg-Langevin equation for the signal field, Eq. (s11), the solution for the signal operator is given by:

$$\hat{a}_s(q_s, \omega_s, 0) = \int d\omega_i \int \frac{d^2q_i}{(2\pi)^2} \int_L^0 dz\, \hat{\mathcal{B}}_s(q_s, q_i, \omega_i, z)$$
$$+ \int_L^0 dz \int_L^0 \tilde{\beta}(q_s, \omega_s)\, N(\omega_s)\hat{f}_s(q_s, \omega_s, z)e^{ik_{s,z}(q_s,\omega_s)z}, \quad (s17.1)$$

where the kernel $\hat{\mathcal{B}}_s(q_s, q_i, \rho, \omega_i)$ is given by:

$$\hat{\mathcal{B}}_s(q_s, q_i, \omega_i, z) =$$
$$= \tilde{\beta}(q_s, \omega_s)\tilde{\sigma}(\omega_s, \omega_i)\hat{E}_{i,z}^\dagger(q_i, \omega_i, z)e^{iz(k_{p,z}(q_p,\omega_{p,0})+k_{s,z}(q_s,\omega_s)-G)} \times \quad (s17.2)$$
$$\times \delta(\omega_{p,0} - \omega_i - \omega_s)\delta(\vec{q}_s + \vec{q}_i - \vec{q}_p),$$

and we denoted $\tilde{\beta}(q_s, \omega_s) \equiv \frac{\omega_s\mu_0}{2k_{s,z}(q_s,\omega_s)\beta(q_s,\omega_s)}$ and $\tilde{\sigma}(\omega_s, \omega_i) \equiv -|\vec{G}||\vec{E}_{p,0}|\omega_s\sigma(\omega_s, \omega_i)(\hat{e}_s \cdot \hat{e}_p)$.

The z component of the vector field operator $\hat{E}_i$ in Eq. (s17.2) arises from the fact that $\vec{G}$ points in the -z direction, so the term $\hat{E}_i^\dagger(q_i, \omega_i, z)(\hat{e}_i^* \cdot \vec{G})$ in the (operator-valued) nonlinear current, Eq. (s5), is equal to $-|\vec{G}|\hat{E}_{i,z}^\dagger(q_i, \omega_i, z)$.

By Fourier transforming Eq. (s17) we obtain:

$$\hat{a}_s(\rho_f, t, 0) = \quad (s18)$$
$$= \frac{1}{(2\pi)^3}\int d^2q_s\, e^{i\vec{q}_s\cdot\vec{\rho}_f}\int d\omega_s \left(\int d\omega_i \int \frac{d^2q_i}{(2\pi)^2}\int_L^0 dz\, \hat{\mathcal{B}}_s(q_s, q_i, \omega_i, \omega_s, z)e^{-i\omega_s t} + \right.$$
$$\left. \int_L^0 dz \int_L^0 \tilde{\beta}(q_s, \omega_s)\, N(\omega_s)\hat{f}_s(q_s, \omega_s, z)e^{ik_{s,z}(q_s,\omega_s)z}\right) + c.c.\,.$$

Next, we substitute the above equation into Eq. (1) of the main text to calculate the count rate.

$$\Gamma_s = \int_0^\infty d\omega_s' \frac{e^{i\omega_s' t}}{(2\pi)^3} \int d^2 q_s' \, e^{-i\vec{q}_s' \cdot \vec{\rho_f}} \int d\omega_i' \int \frac{d^2 q_i'}{(2\pi)^2} \int_L^0 dz' \, \langle vac|\widehat{\mathcal{B}}_s^\dagger(q_s', q_i', \omega_i', \omega_s', z') \quad (s19)$$

$$\times \int_0^\infty d\omega_s \frac{e^{-i\omega_s t}}{(2\pi)^3} \int d^2 q_s \, e^{i\vec{q}_s \cdot \vec{\rho_f}} \int d\omega_i \int \frac{d^2 q_i}{(2\pi)^2} \int_L^0 dz \, \widehat{\mathcal{B}}_s(q_s, q_i, \omega_i, \omega_s, z)|vac\rangle.$$

Note that since the eigen value of $\hat{f}_s|0_s\rangle$ is zero, the noise current in the signal mode does not directly contribute to the count rate. However, this does not imply that absorption of the signal photons has no effect on the count rate. It affects the count rate indirectly through the $e^{ik_{s,z}z}$ term in $\widehat{\mathcal{B}}_s$, as $k_{s,z}$ is complex.

After some calculation, we find:

$$\Gamma_s = \quad (s20.1)$$

$$\frac{1}{(2\pi)^3} \int d^2 q_s' \, \delta(\vec{q}_s' + \vec{q}_i' - \vec{q}_p) e^{-i\vec{q}_s' \cdot \vec{\rho_f}} \int d\omega_i' \, e^{-i\omega_i' t} \int \frac{d^2 q_i'}{(2\pi)^2} \int_L^0 dz' \, \langle vac|\widehat{M}_s^\dagger(q_s', q_i', \omega_i', z')$$

$$\times \frac{1}{(2\pi)^3} \int d^2 q_s \, \delta(\vec{q}_s + \vec{q}_i - \vec{q}_p) e^{i\vec{q}_s \cdot \vec{\rho_f}} \int d\omega_i \, e^{i\omega_i t} \int \frac{d^2 q_i}{(2\pi)^2} \int_L^0 dz \, \widehat{M}_s(q_s, q_i, \omega_i, z)|vac\rangle,$$

where the operator $\widehat{M}_s$ is given by:

$$\widehat{M}_s(q_s, q_i, \omega_i, z) = \tilde{\beta}(q_s, \omega_s)\tilde{\sigma}(\omega_s, \omega_i)\hat{E}_{i,z}^\dagger(q_i, \omega_i, z)e^{iz(k_{p,z}(q_p,\omega_{p,0})+k_{s,z}(q_s,\omega_s)-G)} \quad (s20.2)$$

with $\omega_s$ given by: $\omega_s = \omega_{p,0} - \omega_i$.

Using Eq. (s16) for the vacuum fluctuations of the idler electric field, Eq. (s20.1) reduces to:

$$\Gamma_s = \quad (s21.1)$$

$$\int_L^0 dz' \int_L^0 dz \frac{1}{(2\pi)^3} \int d^2 q_s \int d\omega_i \int \frac{d^2 q_i}{(2\pi)^2} \frac{\hbar \omega_i^2}{\pi \epsilon_0 c^2} \text{Im} \, g_{q_i}^{(z,z)}(z', z, \omega_i) \times$$

$$\frac{1}{(2\pi)^3} \int d^2 q_s' \int d\omega_i' \int \frac{d^2 q_i'}{(2\pi)^2} \delta(\vec{q}_s' + \vec{q}_i' - \vec{q}_p)\delta(\vec{q}_s + \vec{q}_i - \vec{q}_p)\delta(q_i - q_i')\delta(\omega_i - \omega_i')e^{i\vec{\rho_f}\cdot(\vec{q}_s - \vec{q}_s')}e^{it(\omega_i - \omega_i')}Y_s^*(q_s', \omega_i', z')Y_s(q_s, \omega_i, z),$$

where the function $Y_s$ is given by:

$$Y_s(q_s, \omega_i, z) = \tilde{\beta}(q_s, \omega_s)\tilde{\sigma}(\omega_s, \omega_i)e^{iz(k_{p,z}(q_p,\omega_{p,0})+k_{s,z}(q_s,\omega_s)-G)} \quad (s21.2)$$

Using the properties of the delta function, we obtain:

$$\Gamma_s = \int_L^0 dz' \int_L^0 dz \int d\omega_i \int \frac{d^2 q_s}{(2\pi)^2} \frac{\hbar \omega_i^2}{\pi \epsilon_0 c^2} Y_s^*(q_s, \omega_i', z')Y_s(q_s, \omega_i, z)\text{Im} \, g_{q_i}^{(z,z)}(z', z, \omega_i) \quad (s22.1)$$

with:

$$\vec{q}_i = \vec{q}_p - \vec{q}_s \qquad (s22.2)$$

According to Eq. (s21.2), the term $\Upsilon_s^*(q_s, \omega_i, z')\Upsilon_s(q_s, \omega_i, z)$ is given by:

$$\Upsilon_s^*(q_s, \omega_i', z')\Upsilon_s(q_s, \omega_i, z) = |C(q_s, q_p, \omega_i)|^2 e^{i\Delta k_{p,s}(z-z')}, \qquad (s23.1)$$

where:

$$C(q_s, q_p, \omega_i) = \tilde{\beta}(q_s, \omega_s)\tilde{\sigma}(\omega_s, \omega_i), \qquad (s23.2)$$

and

$$\Delta k_{p,s} = k_{p,z}(q_p, \omega_{p,0}) + k_{s,z}(q_s, \omega_{p,0} - \omega_i) - G. \qquad (s23.3)$$

Therefore, the count rate is given by:

$$\Gamma_s = \int d\omega_i \int \frac{d^2 q_s}{(2\pi)^2} \frac{\hbar \omega_i^2}{\pi \epsilon_0 c^2} |C(q_s, q_p, \omega_i)|^2 \frac{q_i^2}{k_i^2(\omega_i)} \\ \times \int_L^0 dz' \int_L^0 dz\, e^{i\Delta k_{p,s}(z-z')} \text{Im}\{g_{q_i}^{(z,z)}(z', z, \omega_i)\}. \qquad (s24)$$

**Eq. (s24) is the signal's count rate equation presented in the main text**.

To evaluate this expression, we need to compute the imaginary part of $g_{q_i(z,z)}(z', z, \omega_i)$, which is the Green's function solution for Eq. (s13) given by[6,8]:

$$g_{q_i}^{(z,z)}(z', z, \omega_i) = \qquad (s25)$$

$$\overbrace{\frac{i}{2k_{i,z}(\omega_i, q_i)} \frac{q_i^2}{k_i^2(\omega_i)} e^{ik_{i,z}(\omega_i, q_i)\cdot|z-z'|}}^{g_{q_i(free)}^{(z,z)}(z',z,\omega_i)} + \overbrace{\frac{i}{2k_{i,z}(\omega_i, q_i)} \frac{q_i^2}{k_i^2(\omega_i)} r_p(\omega_i, q_i) e^{ik_{i,z}(\omega_i, q_i)\cdot(z+z')}}^{g^{(z,z)}_{i(ref)}(z',z,\omega_i)},$$

where $r_p(\omega_i, q_i)$ is the Fresnel reflection coefficient for a "p" polarized wave and for the case of a metal-air interface $r_p$ it is given by:

$$r_p(\omega_i, q_i) = \frac{k_{i,z}^{(m)}(\omega_i, q_i) - \varepsilon_m(\omega_i)k_{i,z}^{(a)}(\omega_i, q_i)}{k_{i,z}^{(m)}(\omega_i, q_i) + \varepsilon_m(\omega_i)k_{i,z}^{(a)}(\omega_i, q_i)}. \qquad (s26)$$

The indices "m" and "a" represent metal and air, respectively. The z component of k vectors in the metal and air are given by $k_{i,z}^{(l=m,a)}(\omega_i, q_i) = \sqrt{k_{i,l}^2(\omega_i) - q_i^2}$.

This Green's function, $g_{q_i}$, comprises a freely propagating wave in an infinite medium with metal permittivity and a wave reflected at the metal-air interface. The reflected field incorporates Fresnel's reflection coefficient of only a 'p' polarized electric field. Since surface plasmons are only excited by 'p' polarized electric field, it is essential that the idler electric field possesses a component perpendicular to the plane of incidence. This is ensured by the nonlinear current, Eq. (s5), which is proportional to $\hat{e}_i^* \cdot \vec{G}$.

$g_{q_i}^{(z,z)}$ also depends on the complex longitudinal idler's wave-vector, $k_{i,z}$ which is taken to lie in the upper half of the complex plane (that is, we choose $\text{Im}(k_{i,z}) > 0$).

Inserting $g_{q_i}^{(z,z)}$ given in Eq. (s25) into Eq. (s24) we get:

$$\Gamma_s = \Gamma_{s,free} + \Gamma_{s,ref}, \qquad (s27)$$

where $\Gamma_{s,ref}$ and $\Gamma_{s,free}$ are the contributions to the count rate from the reflected and free propagation Green's functions, respectively.

$\Gamma_{s,free}$ is given by:

$$\Gamma_{s,free} = \int d\omega_i \int \frac{d^2 q_s}{(2\pi)^2} \frac{\hbar \omega_i^2}{\pi \epsilon_0 c^2} |C(q_s, q_p, \omega_i)|^2 \frac{q_i^2}{k_i^2(\omega_i)} \Pi_{free}(q_s, q_p, \omega_i) \quad (s28.1)$$

where $\Pi_{free}$ is defined as:

$$\Pi_{free} = \int_L^0 dz' \int_L^0 dz\, e^{i\Delta k_{p,s}(z-z')} \text{Im}\left\{\frac{i}{2k_{i,z}(\omega_i, q_i)} \frac{q_i^2}{k_i^2(\omega_i)} e^{ik_{i,z}(\omega_i, q_i)\cdot |z-z'|}\right\}, \quad (s28.2)$$

And $\Gamma_{s,ref}$ is given by:

$$\Gamma_{s,ref} = \int d\omega_i \int \frac{d^2 q_s}{(2\pi)^2} |C(q_s, q_p, \omega_i)|^2 \frac{q_i^2}{k_i^2(\omega_i)} \Pi_{ref}(q_s, q_p, \omega_i) \qquad (s28.3)$$

where $\Pi_{ref}$ is defined as:

$$\Pi_{ref} = \int_L^0 dz' \int_L^0 dz\, \frac{\hbar \omega_i^2}{\pi \epsilon_0 c^2} e^{i\Delta k_{p,s}(z-z')} \text{Im}\left\{\frac{i}{2k_{i,z}(\omega_i, q_i)} r_p(\omega_i, q_i) e^{ik_{i,z}(\omega_i, q_i)\cdot (z+z')}\right\}. \quad (s28.4)$$

To compute $\Gamma_{s,free}$ and $\Gamma_{s,ref}$ we first need to evaluate $\Pi_{free}$ and $\Pi_{ref}$.

**Evaluation of $\Pi_{free}$:**

First, we use the following mathematical identity[8,9]:

$$\frac{i}{2k_{i,z}} e^{ik_{i,z}|z-z'|} = \frac{1}{2\pi} \int_{-\infty}^{\infty} dk\, \frac{e^{ik(z-z')}}{k^2 - k_{i,z}^2}. \qquad (s29)$$

The poles of the function $\frac{e^{ik(z-z')}}{k^2-k_{i,z}^2}$ are located at $k = k_{i,z}$ and $k = -k_{i,z}$. Recall that we choose $k_{i,z}$ in the upper half of the complex plane (i.e. $\text{Im}\{k_{i,z}\} > 0$), which places $-k_{i,z}$ in the lower half of the complex plane.

To calculate the residue of $\frac{e^{ik(z-z')}}{k^2-k_{i,z}^2}$ at $k = k_{i,z}$ we select a contour that starts at $\text{Re}(k) = -\infty$, continuous to $\text{Re}(k) = +\infty$ and returns to the starting point in an anti-clockwise arch, ensuring that $\frac{e^{ik(z-z')}}{k^2-k_{i,z}^2}$ vanishes at $|k| \to \infty$. For the residue at $k = k_{i,z}$ we use the same procedure but with a clockwise contour.

Using the residue theorem, and noting that clockwise and counterclockwise contours introduce pre-factors of $(-2\pi i)$ and $2\pi i$, respectively, we obtain:

$$\frac{1}{2\pi}\int_{-\infty}^{\infty} dk \frac{e^{ik(z-z')}}{k^2 - k_{i,z}^2} = \frac{1}{2\pi}\left(2\pi i \times \text{Res}\left\{\frac{e^{ik(z-z')}}{k^2 - k_{i,z}^2}\right\}_{k=k_{i,z}} + (-2\pi i) \times \text{Res}\left\{\frac{e^{ik(z-z')}}{k^2 - k_{i,z}^2}\right\}_{k=-k_{i,z}}\right) \quad (s30)$$

$$= i\left(\frac{e^{ik_{i,z}(z-z')}}{2k_{i,z}} + \frac{e^{-ik_{i,z}(z-z')}}{2k_{i,z}}\right).$$

Noting that: $\text{Im}\left\{\frac{1}{2\pi}\int_{-\infty}^{\infty} dk \frac{e^{ik(z-z')}}{k^2-k_{i,z}^2}\right\} = \text{Re}\left\{\frac{e^{ik_{i,z}(z-z')}}{2k_{i,z}} + \frac{e^{-ik_{i,z}(z-z')}}{2k_{i,z}}\right\},$

we obtain

$$\Pi_{free} = \int_L^0 dz' \int_L^0 dz\, e^{i\Delta k_{p,s}(z-z')}\left(\text{Re}\left\{\frac{e^{ik_{i,z}(z-z')}}{2k_{i,z}}\right\} + \text{Re}\left\{\frac{e^{-ik_{i,z}(z-z')}}{2k_{i,z}}\right\}\right). \quad (s31)$$

Then, using the relation:

$$\text{Re}\left\{\frac{e^{\pm ik_{i,z}(z-z')}}{2k_{i,z}}\right\}$$

$$= \frac{1}{2}\left(e^{i(\pm\text{Re}(k_{i,z})+i\text{Im}(k_{i,z}))(z-z')}\frac{1}{2k_{i,z}} + e^{i(\mp\text{Re}(k_{i,z})+i\text{Im}(k_{i,z}))(z-z')}\left(\frac{1}{2k_{i,z}}\right)^*\right),$$

and noting that $2\text{Re}\left\{\frac{1}{2k_{i,z}}\right\} = \frac{1}{2k_{i,z}} + \left(\frac{1}{2k_{i,z}}\right)^*$, we integrate over $z'$ and $z$ and obtain the result:

$$\Pi_{free} = \text{Re}\left\{\frac{1}{2k_{i,z}}\right\}(|a_+|^2 + |a_-|^2), \quad (s32.1)$$

where:

$$a_+ = \int_L^0 dz\, e^{iz(\text{Re}(\Delta k_{p,s}+k_{i,z})+i\text{Im}(k_{i,z}+\Delta k_{p,s}))} = \frac{1 - e^{i(\text{Re}(\Delta k_{p,s}+k_{i,z}))L - \text{Im}(\Delta k_{p,s}+k_{i,z})L}}{i(\text{Re}(\Delta k_{p,s}+k_{i,z})) - \text{Im}(\Delta k_{p,s}+k_{i,z})}, (s32.2)$$

and

$$a_- = \int_L^0 dz\, e^{-iz(\text{Re}(\Delta k_{p,s}-k_{i,z})-i\text{Im}(k_{i,z}+\Delta k_{p,s}))} = \frac{1-e^{-i(\text{Re}(\Delta k_{p,s}-k_{i,z}))L-\text{Im}(\Delta k_{p,s}+k_{i,z})L}}{-i(\text{Re}(\Delta k_{p,s}-k_{i,z}))-\text{Im}(\Delta k_{p,s}+k_{i,z})}. (s32.3)$$

Note that the oscillatory components in $a_+$ and $a_-$, the exponents $e^{-i(\text{Re}(\Delta k_{p,s}+k_{i,z}))L}$ and $e^{-i(\text{Re}(\Delta k_{p,s}-k_{i,z}))L}$, respectively, describe idler photons propagating in opposite directions along the z axis, thereby accounting for both possibilities.

**Evaluation of $\Pi_{ref}$:**

To evaluate $\Pi_{ref}$, we use the following two identities:

$$\text{Im}\left\{\frac{i}{2k_{i,z}} r_p e^{ik_{i,z}(z+z')}\right\}$$
$$= \frac{1}{2i}\left(i\frac{r_p}{2k_{i,z}} e^{i(\text{Re}(k_{i,z})+i\text{Im}(k_{i,z}))(z+z')} - (-i)\left(\frac{r_p}{2k_{i,z}}\right)^* e^{-i(\text{Re}(k_{i,z})-i\text{Im}(k_{i,z}))(z+z')}\right), \quad (s33.1)$$

and:

$$\int_L^0 dz\, e^{iz\left(\text{Re}(\Delta k_{p,s}) \pm \text{Re}(k_{i,z}) + i\left(\text{Im}(k_{i,z}) + \text{Im}(\Delta k_{p,s})\right)\right)}$$
$$= \left(\int_L^0 dz'\, e^{-iz'\left(\text{Re}(\Delta k_{p,s}) \pm \text{Re}(k_{i,z}) - i\left(\text{Im}(k_{i,z}) + \text{Im}(\Delta k_{p,s})\right)\right)}\right)^*. \quad (s33.2)$$

To obtain:

$$\Pi_{ref} = \text{Re}\left(\frac{r_p}{2k_{i,z}} a_+ a_-\right). \quad (s33)$$

**The final expression for the signal count rate is therefore given by the following compact formula:**

$$\Gamma_s = \int d\omega_i \int \frac{d^2 q_s}{(2\pi)^2} \frac{\hbar \omega_i^2}{\pi \epsilon_0 c^2} |C(q_s, q_p, \omega_i)|^2 \frac{q_i^2}{k_i^2(\omega_i)}$$
$$\times \left(\overbrace{\text{Re}\left(\frac{r_p(\omega_i, q_i)}{2k_{i,z}(\omega_i, q_i)} a_+ a_-\right)}^{reflected} + \overbrace{\text{Re}\left\{\frac{1}{2k_{i,z}(\omega_i, q_i)}\right\} (|a_+|^2 + |a_-|^2)}^{free}\right), \quad (s34)$$

with: $\vec{q}_i = \vec{q}_p - \vec{q}_s$.

# I.  Extended experimental data

To provide supporting evidence that the resonance enhancement observed in aluminum in the present work is not a result of the experimental setup, we present the signal spectrum of a diamond crystal measured with the same setup. These data were collected during a previous experiment at the GALAXIES beamline [10].

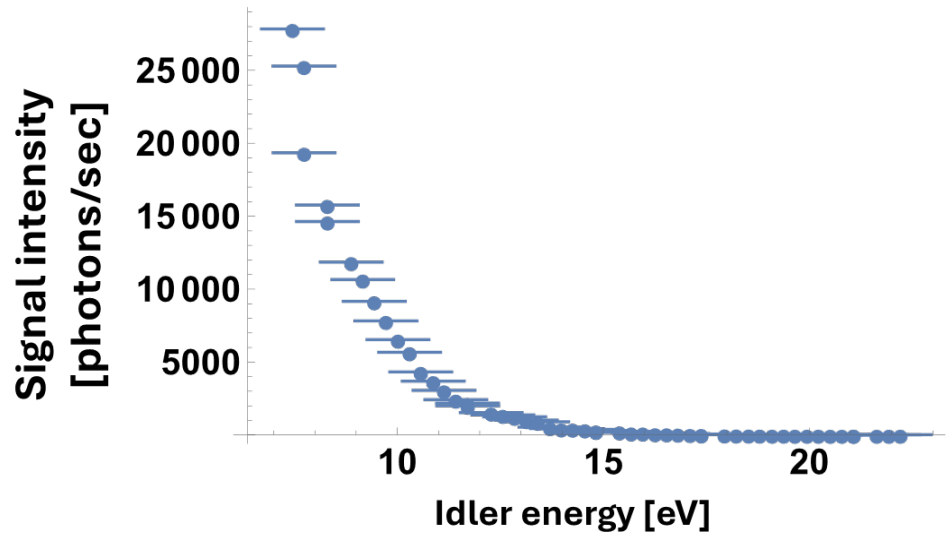

**Fig. S1| The measured signal spectrum of a diamond crystal for idler energies between 7 eV and 22 eV was obtained using the same setup as in the present work.** As shown, no enhancement is observed near the surface plasmon resonance energy (~10.65 eV).

# References

1. Freund, I. & Levine, B. F. Optically Modulated X-Ray Diffraction. *Phys Rev Lett* 25, 1241–1245 (1970).

2. Jha, S. S. & Woo, J. W. F. Nonlinear response of electrons in a solid to X-rays. Il Nuovo Cimento B 10, 229–239 (1972).

3. Glover, T. E. *et al.* X-ray and optical wave mixing. *Nature 2012 488:7413* 488, 603–608 (2012).

4. A. Schori1, C. Bömer, D. Borodin, S. P. Collins, B. Detlefs, M. Moretti Sala, S. Yudovich, and S. Shwartz. Parametric Down-Conversion of X Rays into the Optical Regime. Phys Rev Lett 119, (2017).

5. Shwartz, S. *et al.* X-ray parametric down-conversion in the Langevin regime. *Phys Rev Lett* 109, (2012).

6. Scheel, S. & Buhmann, S. Y. Macroscopic quantum electrodynamics-concepts and applications. *Acta Physica Slovaca* 58, 675–809 (2008).

7. Peřina, Jan. Coherence and statistics of photons and atoms. 520 (2001).

8. Novotny, L. & Hecht, B. Principles of Nano-Optics. *Principles of Nano-Optics* 9781107005464, 1–564 (2012).

9. Gruner, T. & Welsch, D. G. Green-function approach to the radiation-field quantization for homogeneous and inhomogeneous Kramers-Kronig dielectrics. *Phys Rev A (Coll Park)* 53, 1818–1829 (1996).

10. Borodin, D., Schori, A., Rueff, J. P., Ablett, J. M. & Shwartz, S. Evidence for Collective Nonlinear Interactions in X Ray into Ultraviolet Parametric Down-Conversion. *Phys Rev Lett* 122, (2019).